\documentclass[twocolumn]{autart}    
\usepackage{amsmath,amssymb}
\usepackage{bm}

\newtheorem{hyp}{\bf Assumption}
\newtheorem{example}{\bf Example}
\newtheorem{lemma}{\bf Lemma}
\newtheorem{corollary}{\bf Corollary}
\newtheorem{proposition}{\bf Proposition}
\usepackage{tikz}
\usetikzlibrary{arrows,shapes,calc}
\usepackage{pgfplots}
\usetikzlibrary{positioning} 

\tikzstyle{solid}=                   [dash pattern=]
\tikzstyle{dotted}=                  [dash pattern=on \pgflinewidth off 2pt]
\tikzstyle{densely dotted}=          [dash pattern=on \pgflinewidth off 1pt]
\tikzstyle{loosely dotted}=          [dash pattern=on \pgflinewidth off 4pt]
\tikzstyle{dashed}=                  [dash pattern=on 3pt off 3pt]
\tikzstyle{densely dashed}=          [dash pattern=on 3pt off 2pt]
\tikzstyle{loosely dashed}=          [dash pattern=on 3pt off 6pt]
\tikzstyle{dashdotted}=              [dash pattern=on 3pt off 2pt on \the\pgflinewidth off 2pt]
\tikzstyle{densely dashdotted}=      [dash pattern=on 3pt off 1pt on \the\pgflinewidth off 1pt]
\tikzstyle{loosely dashdotted}=      [dash pattern=on 3pt off 4pt on \the\pgflinewidth off 4pt]

\usepackage{graphicx}      
\begin{document}

\begin{frontmatter}

\title{A New Contraction-Based NMPC Formulation Without Stability-Related Terminal Constraints} 


\author[Mazen]{Mazen Alamir}\ead{mazen.alamir@grenoble-inp.fr},    

\address[Mazen]{CNRS/University of Grenoble Alpes, France. }  

\begin{keyword}                           
Model Predictive Control, Nonlinear Systems, Short Prediction Horizon.             
\end{keyword}                             

\begin{abstract}                          
Contraction-Based Nonlinear Model Predictive Control (NMPC) formulations are attractive because of the generally short prediction horizons they require and the needless use of terminal set computation that are commonly necessary to guarantee stability. However, the inclusion of the contraction constraint in the definition of the underlying optimization problem often leads to non standard features such as the need for multi-step open-loop application of control sequences or the use of multi-step memorization of the contraction level that may induce unfeasibility in presence of unexpected disturbance. This paper proposes a new formulation of contraction-based NMPC in which no contraction constraint is explicitly involved. Convergence of the resulting closed-loop behavior is proved under mild assumptions. An illustrative example is proposed in order to assess the relevance of the proposed formulation. 
\end{abstract}

\end{frontmatter}

\section{Introduction}
Provable closed-loop stability in the majority of NMPC formulations results from the use of terminal constraints on the state. In the early formulations \cite{Keerthi1988,Mayne1990}, stringent equality constraint on the state is used. Then relaxations were introduced through the combined use of terminal set inclusion and appropriate terminal penalty. The many different ways to choose these two items were unified in \cite{Mayne2000} where it has been shown that the terminal set should be controlled-invariant under some {\em local}  feedback control that makes the terminal penalty a control-Lyapunov function. This pair of terminal set and terminal penalty function are the most often computed based on Linear Quadratic Regulator (LQR) design if the linearized system around the targeted state is stablilizable. Otherwise, invariant sets computational machinery can be used for nonlinear systems \cite{Blanchini19991747,Limon2003,Rakovic2005}. In particular, the recently proposed scheme \cite{Lazar2015} can be used for purely nonlinear systems through extensive use of the finite step Lyapunov function paradigm. Contractive NMPC schemes based on interval analysis \cite{Jaulin2001} have been also proposed \cite{Wan2004,Wan2007} an applied to relatively simple 2D robotic examples where this technique is still usable. \ \\ \ \\ Regardless of the way the pair of terminal state and terminal penalty is computed, the feasibility of the associated terminal constraint generally needs long prediction horizons to be used in the MPC formulation. Moreover, the presence of this constraint makes the computation of the optimal solution a difficult task. This may explain why many practitioners {\em confess} never including such stability-related constraints in their formulations even in applications where the latter is almost dedicated to stabilization.   \ \\ \ \\ 
On the other hand, it has been shown quite early \cite{Alamir1995} that provable stability can be obtained without terminal stability-related constraint by using {\em sufficiently long} prediction horizon \cite{Grim:2005,Jadbabaie:2005}. More recent results followed, [see \cite{Grune2010,Grune2011,Boccia:2014} and the references therein] where deeper analysis is obtained regarding this fact. However, the underlying argument remained that with sufficiently long prediction horizon, the optimal decisions necessarily lead to open-loop trajectories with terminal appropriate properties.  \ \\ \ \\ For obvious computational reasons, one might be  interested in formulations involving short prediction horizons and no stability-related terminal constraints. Following the previous argumentation, this might appear paradoxical. This is precisely where contractive formulations enter into picture. Indeed, the contraction property for a controlled system is the systematic ability to find a control sequence $\bm u$ that steers the state of the system from its current value $x_k$ to a new state $x_{k+N}$ where the value of some positive definite function $W$ is contracted by some ratio $\gamma\in (0,1)$, namely $W(x_{k+N})\le \gamma W(x_k)$. Now when $N=1$, this property is satisfied only for difficult-to-find standard controlled-Lyapunov functions. However, as $N$ gets a little bit higher, the property becomes true for a wider class of functions, referred to as $N$-step Lyapunov functions \cite{Bobiti2014}. More interestingly, it can be shown \cite{Alamir2006,Bobiti2014} that for stabilizable systems, {\em any} positive definite function $W$ satisfies the contraction property for appropriate $N$. Such $N$'s are generally much shorter than the one that would be needed to make standard terminal constraints feasible for a large set of possible initial states. \ \\ \ \\ 
The difficulty in including the contraction property in the MPC formulation comes from the receding-horizon implementation of the resulting optimal sequence. Indeed, assume that an open-loop {\em contractive} trajectory is found (at instant $k$) such that $W(x_{k+N}^{ol})\le \gamma W(x_k^{ol})$, then it might still be true that 
$\gamma W(x_{k+1}^{ol})>W(x_k^{ol})$ since $W$ is only a finite-step Lyapunov function and hence not monotonically decreasing. This means that if the problem is re-formulated at instant $k+1$ using the constraint 
$W(x_{k+1+N})\le \gamma W(x_{k+1}^{ol})$ then this does not guarantee closed-loop contraction of $W$. This explains why in the earlier use of the contraction property in MPC formulation \cite{Kothare2000}, two possible alternatives were proposed to enhance closed-loop contraction. In the first, the contractive open-loop trajectory is applied in open-loop until contraction occurs. In the second, the contraction level $\gamma W(x_{k}=x_{past})$ is memorized and used in formulating the subsequent optimization problems with the constraint $\min_{i=1,\dots,N}W(x_{k+i})\le \gamma W(x_{past})$ until contraction occurs at some instant $k+i^*$ at which the updating rule $x_{past}=x_{k+i^*}$ is adopted and the process is repeated. These two alternatives are obviously not satisfactory since in the former, the system is left in open-loop while in the second, the use of memorized level might lead to unfeasibility problem in the presence of disturbance.
These drawbacks motivated the contractive scheme proposed in \cite{alamir2007low} where no stability-related constraint is used in the MPC formulation.\ \\ \ \\  
The present paper improves the formulation proposed in \cite{alamir2007low} by using standard cost function together with a stability-dedicated penalty term while in \cite{alamir2007low}, only the contractive function is used in the cost function which makes the formulation of \cite{alamir2007low} exclusively dedicated to stabilization. Moreover, state constraints are considered while \cite{alamir2007low} considered only control saturation. \ \\ \ \\ 
This paper is organized as follows: First, the definitions and notation used throughout the paper are introduced in Section \ref{secdefnot} which also introduces the Assumptions needed to derive the main result. Section \ref{secproposed} introduces the proposed contractive MPC formulation together with the main convergence results. Section \ref{secimplementation} proposes a modified formulation to address some computational issues. An illustrative example is given in Section \ref{secexample} while Section \ref{secconc} summarizes the contribution and gives hints for further investigation.

\section{Definitions and notation} \label{secdefnot} 
This paper concerns nonlinear systems of the form:
\begin{equation}
x_{k+1}=f(x_k,u_k) \label{systemmm} 
\end{equation} 
where $x\in \mathbb{R}^{n}$ is the state, $u\in \mathbb{R}^{m}$ is the control input. Given a sequence $\bm{u}:=(u^{(1)},\dots,u^{(N)})\in \mathbb{R}^{m\times N}$ of future control inputs together with some initial state $x$, the resulting state trajectory is denoted by $\bm{x}^{\bm{u}}(x):=(x^{(1)},\dots,x^{(N)})$ where $x^{(1)}=f(x,u^{(1)})$ and $x^{(i+1)}=f(x^{(i)},u^{(i+1)})$. In the sequel, the notation $\bm{u}_\ell=u^{(\ell)}$ and $\bm{x}^{\bm{u}}_\ell(x)=x^{(\ell)}$ is used when needed, namely $\bm{x}^{\bm{u}}_\ell(x)$ is the state reached $\ell$-steps-ahead starting from the initial state $x$ and applying the sequence of controls $\bm{u}_1,\dots,\bm u_\ell$. \ \\ \ \\ 
Regarding the constraints, it is assumed that $u$ belongs to a compact set $\mathbb U\subset \mathbb{R}^{m}$ and that the set of admissible states is given by $\mathbb G:=\left\{x\ \vert \ g(x)\le 0\right\}$. Moreover the following assumption is adopted in the sequel:
\begin{hyp} \label{hyp1} 
$\mathbb G$ is a $\mathbb U$-Controlled-Invariant set that contains a neighborhood of the origin. 
\end{hyp}
Note that this Assumption is generally not satisfied if $g(x)$ simply expresses the simple enumeration of physical constraints on $x$. However, it can be made satisfied by appropriate tightening of the set of constraints as it is shown in the following simple example:
\begin{example} \label{ex1} 
Consider the discrete-time version of the system given by $\ddot r=u$ where $u\in \mathbb U:=[-\bar u,\bar u]$ and $r\in [-\bar r,+\bar r]$. Take $x=(r,\dot r)$. If one uses the trivial definition $g(x)=\vert x_1\vert-\bar r$ then $\mathbb G$ does not satisfy Assumption \ref{hyp1}.  However, if the constraint is tightened so that:
\begin{equation}
g(x):= \begin{pmatrix}
\vert x_1\vert -\bar r\\
x_1+x_2\tau-Sign(x_2)[\frac{1}{2}\bar u\tau^2+\bar r]
\end{pmatrix} 
\end{equation} 
then the resulting $\mathbb G$ meets Assumption \ref{hyp1}. This is because the additional constraint limits the speed $x_2$ so that admissible brake force can avoid the violation of the {\em original} constraint. 
\end{example}
\begin{rem}
Assumption \ref{hyp1} is a viability assumption that is common i n MPC formulation without terminal constraints and/or cost. See for instance similar statement of this assumption in \cite{Kerrigan:2000,Blanchini:2008}. 
\end{rem}
Regarding the contraction property, the following assumption is used:
\begin{hyp} \label{hyp2} 
There exists a positive definite function $W$, a contraction factor $\gamma\in (0,1)$ and a prediction horizon $N$ such that  $\forall x\in \mathbb G$, $\exists\bm u\in \mathbb U^N$ such that:
\begin{align}
&\forall \ell\in \{1,\dots,N\}\qquad \bm x^{\bm u}_\ell(x)\in \mathbb G \label{admissiblility}  \\
&\underline W(x,\bm u,N):=\min_{\ell=1}^N \Bigl[W(\bm x^{\bm u}_\ell(x))\Bigr]\le \gamma W(x) \label{defdecontraction} 
\end{align}
\end{hyp}
\begin{rem} \label{rem0} 
Note that Assumption \ref{hyp2} implicitly implies Assumption \ref{hyp1}. That is the reason why Assumption \ref{hyp1} never appears in the formulation of the results of this contribution. The reason for which Assumption \ref{hyp1} is explicitly stated though is to underline the need for constraints tightening so that the admissible set $\mathbb G$ becomes controlled-invariant.     
\end{rem}
In the sequel, the argument of the minimization problem in (\ref{defdecontraction}) over a prediction horizon of length $N$ is denoted by $\ell_{opt}(x,\bm u,N)$. More generally, given a prediction horizon $q\le N$, the following notation is used:
\begin{equation}
\ell_{opt}(x,\bm u,q):= {\rm arg}\min_{\ell\in \{1,\dots,q\}} W(\bm x_\ell^{\bm u}(x)) 
\end{equation} 
As mentioned in the introduction, finding a pair $(W,N)$ satisfying Assumption \ref{hyp2} is much easier than looking for standard one-step Lyapunov function as any positive definite function $W$ would be successful candidate for {\em $N$} moderately large. For a specific system, such function can even be linked to some physical considerations. In the absence of such facility, randomized optimization procedure \cite{alamo2009} might be used to set the pair $(W,N)$ that satisfies (\ref{defdecontraction}) with extremely high probability. Note also that the {\em machinery} developed in \cite{Grune2010} around the concept of multi-step Lyapunov inequality can lead to more checkable conditions that guarantee the contraction property.  \ \\ \ \\ 
It is assumed that a stage cost function $L(x,u)$ is used to express the control objective. For a control sequence $\bm u$, the following notation is used:
\begin{equation}
\Phi(x,\bm u,q):=\sum_{\ell=1}^qL(\bm x^{\bm u}_\ell(x),\bm u_\ell) \label{defdephiq} 
\end{equation} 
Moreover, the following assumption is used regarding the behavior of $L$ inside the admissible domain 
\begin{hyp}\label{hyp3} 
$\exists\bar L>0$ such that:
\begin{equation}
\forall (x,u)\in \mathbb G\times \mathbb U,\quad 0\le L(x,u)\le \bar L \label{defdeLbar} 
\end{equation} 
Moreover, $Q(x):=L(x,0)$ is a positive definite function of the state and such that $Q(x)\le L(x,u)$ for all $u$.
\end{hyp}
This simply means that $L$ is bounded on the set of admissible pairs $(x,u)$ and contains a positive definite penalty on the state regardless of the control value. In the next section, the proposed MPC formulation is given and the behavior of the resulting closed-loop is analyzed. 
\section{The contractive formulation} \label{secproposed} 
Let us define for any $z>0$ and any state $x\in \mathbb G$, the following optimization problem, denoted by $\mathcal P(x,z)$:
\begin{align}
&\min_{(\bm u,q)} \left[J^{(x,z)}(\bm u,q)\right]:=z\cdot \Phi(x,\bm u,q)+\alpha\underline W(x,\bm u,q) \label{costdef1}\\
& \qquad \mbox{\rm under}\qquad   \bm x^{\bm u}_\ell(x)\in \mathbb G\qquad \forall\ell\in \{1,\dots,q\} \label{costdef2} \\
&\qquad  \mbox{\rm and}\qquad\ \  (\bm u,q)\in \mathbb U^N\times \{1,\dots,N\} \label{costdef3} 
\end{align} 
where $z$ is an internal state of the controller with dynamics defined by (\ref{defdezplus}) hereafter. $q$ is the free-prediction horizon which is considered as a decision variable in the proposed formulation. Note that for a given $q$, only the trajectory over the future interval $[k,k+q]$ is involved in the definition of the cost function (\ref{costdef1}) and the constraints (\ref{costdef2}). The two terms involved in the cost function (\ref{costdef1}) are respectively given by (\ref{defdephiq}) and (\ref{defdecontraction}) and represent respectively the stage cost  over the prediction horizon of length $q$ and the lowest value of $W$ over the same prediction horizon.
\begin{rem} \label{rem1} 
Note that the solution of (\ref{costdef1})-(\ref{costdef3}) involves the integer scalar decision variable $q$. Even if $N$ is expected to be small, this can be harmful for the efficiency of the propsoed formulation. This issue is tackled in section \ref{secimplementation} where a computational procedure is proposed that avoids optimization problems involving integer decision variable. 
\end{rem}
\begin{rem} \label{hyp4} 
It is assumed hereafter that in the case where several solutions exist to (\ref{costdef1})-(\ref{costdef3}) with different candidate prediction horizons, then the one with the shortest prediction horizon is selected by the optimizer
\end{rem}
Let us denote by $\bm u^*(x,z)$ and $q^*(x,z)$ the optimal solutions (if any) of the optimization problem (\ref{costdef1})-(\ref{costdef3}) . Moreover, the corresponding value of $\underline W$, $\Phi$, $J$ and $\ell_{opt}$ are denoted by:
\begin{align}
\underline W^*(x,z)&:=\underline W(x,\bm u^*(x,z),q^*(x,z)) \label{defdeWunderstar} \\
\Phi^*(x,z)&:=\Phi(x,\bm u^*(x,z),q^*(x,z)) \label{defdePhistar} \\
J^*(x,z)&:=J^{(x,z)}(\bm u^*(x,z),q^*(x,z))) \label{defdeJstar} \\
\ell^*(x,z)&:=\ell_{opt}(x,\bm u^*(x,z),q^*(x,z))
\end{align} 
The dynamic of the controller's internal state $z$ is given by:
\begin{align}
z_{k+1}=h(x_k,z_k):=\left\{ 
\begin{array}{ll}
 z_k& \mbox{\rm if $W(x_k)>z_k$}  \\
\beta z_k& \mbox{if $W(x_k)\le z_k$}
\end{array}
\right. \label{defdezplus} 
\end{align} 
where $\beta\in (0,1)$ is some fixed constant that can be viewed as a parameter of the controller. \ \\ \ \\ 
This completely defines the MPC feedback by:
\begin{align}
&z^+=h(x,z)\\
&K_{MPC}(x,z):=\bm u^*_1(x,z)
\end{align} 
In what follows, some preliminary results are derived which are used later in the proof of the main result.\ \\ \ \\ 
The first result and its Corollary \ref{cor1} give an explicit and computable upper bound on the optimal cost:
\begin{lemma} \label{lem1} 
{\bf If} Assumptions \ref{hyp2}, \ref{hyp3} are satisfied {\bf then} $\forall (x,z)\in \mathbb G\times \mathbb R_+$, $\mathcal P(x,z)$ is feasible. Moreover:
\begin{equation}
J^*(x,z)\le zN\bar L+\alpha \gamma W(x) \label{vilate} 
\end{equation} 
and the minimum value of the contractive map $W$ is obtained at the end of the trajectory, namely:
\begin{equation}
\ell^*(x,z)=q^*(x,z) \label{ellq} 
\end{equation} 
\end{lemma}
{\sc Proof}. Feasibility is a direct consequence of Assumption \ref{hyp2} since it guarantees the feasibility of (\ref{admissiblility}) which is the only constraints (\ref{costdef2}) that may lead to unfeasibility. Moreover, taking any $\bm u$ satisfying the conditions of Assumption \ref{hyp2}, the corresponding stage cost is obviously lower than $N\bar L$ thanks to (\ref{defdeLbar}) of Assumption \ref{hyp3}  while the second term $\underline W(x,\bm u,N)$ is lower than $\gamma W(x)$ by virtue of (\ref{defdecontraction}). This proves (\ref{vilate}). As for (\ref{ellq}), it can be proved by contradiction. Indeed, if $q^*(x,z)>\ell^*(x,z)$, then the candidate solution $(\bm u^c,q^c):=(\bm u^*(x,z),\ell^*(x,z))$ would correspond to a cost function value satisfying 
\begin{align}
J^{(x,z)}(\bm u^c,q^c)&=J^*(x,z)-z\left[\sum_{k=\ell^*(x,z)+1}^{q^*(x,z)}L(\bm x^{\bm u^*}_k,0)\right]\nonumber \\
&-\alpha\left[q^*(x,z)-\ell^*(x,z)\right]\underline W^*(x,z)< J^*(x,z) \nonumber 
\end{align}  
which contradicts the optimality of $q^*(x,z)$. $\hfill \Box$ 
\begin{corollary} \label{cor1} 
Under Assumptions \ref{hyp2} and \ref{hyp3}, {\bf if} the following conditions hold
\begin{enumerate}
\item $W(x)>z$ and 
\item $\alpha\ge  2N\bar L/(1-\gamma)$
\end{enumerate}  
{\bf then}  the optimal solution satisfies the inequality:
\begin{equation}
J^*(x,z)\le \left[\dfrac{1+\gamma}{2}\right]\alpha W(x) \label{nbnbnhjgkh} 
\end{equation} 
\end{corollary}
{\sc Proof}. Since the condition of Lemma \ref{lem1} are satisfied, inequality (\ref{vilate}) holds, namely 
\begin{equation}
J^*(x,z)\le zN\bar L+\alpha \gamma W(x)
\end{equation} 
and since $z\le W(x)$, one can write:
\begin{equation}
J^*(x,z)\le (N\bar L+\alpha\gamma) W(x)
\end{equation} 
and using the assumption $\alpha\ge  2N\bar L/(1-\gamma)$, the last inequality becomes:
\begin{equation}
J^*(x,z)\le (\dfrac{(1-\gamma)\alpha}{2}+\alpha\gamma) W(x)
\end{equation} 
which gives (\ref{nbnbnhjgkh}) after straightforward manipulation. $\hfill \Box$ \ \\ \ \\ 
While the two preceding results determine bounds on the optimal cost, the following two lemmas characterize the behavior of two successive values of the optimal cost at instants $k$ where $W(x_k)$ is still greater than $z_k$. Lemma \ref{lem2} characterizes this behavior in the case where $q^*(x_k,z_k)>1$ while Lemma \ref{lem3} gives this characterization when $q^*(x_k,z_k)=1$:  
\begin{lemma} \label{lem2} 
Under Assumptions \ref{hyp2} and \ref{hyp3},  {\bf if} the following conditions hold
\begin{enumerate}
\item $q^*(x_k,z_k)>1$ and 
\item $W(x_k)>z_k$
\end{enumerate} 
 {\bf then}  the following inequality holds:
\begin{equation}
J^*(x_{k+1},z_{k+1})\le J^*(x_k,z_k)-z_{k}Q(x_{k+1}) \label{gftrG} 
\end{equation}  
where $Q(\cdot)$ is the positive definite function invoked in Assumption \ref{hyp3}. 
\end{lemma}
{\sc Proof}. Since $q^*(x_k,z_k)>1$, an admissible pair $(\bm u^+,q^+)$ for the optimization problem $\mathcal P(x_{k+1},z_{k+1})$ can be given by:
\begin{align}
q^+&=q^*(x_k,z_k)-1\nonumber \\
\bm u^+&:=\begin{pmatrix}
\bm u^*_2(x_k,z_k) & \dots& \bm u^*_{q^*(x_k,z_k)}(x_k,z_k)
\end{pmatrix} \nonumber 
\end{align} 
But since in this case $z_{k+1}=z_k$ [see (\ref{defdezplus})] and since the terminal cost is unchanged by virtue of (\ref{ellq}) of Lemma \ref{lem1}, the cost of this candidate pair is obviously given by:
\begin{equation*}
J^{(x_{k+1},z_{k+1})}(\bm u^+,q^+)=J^*(x_k,z_k)-z_kL(x_{k+1},\bm u_1^*(x_k,z_k))
\end{equation*} 
and using the fact that $L(x,u)\ge L(x,0)=:Q(x)$ (Assumption \ref{hyp3}), the last inequality gives:
\begin{equation}
J^{(x_{k+1},z_{k+1})}(\bm u^+,q^+)=J^*(x_k,z_k)-z_kQ(x_{k+1})
\end{equation}   
which obviously leads to (\ref{gftrG}) by the very definition of optimality. $\hfill \Box$
\begin{lemma} \label{lem3} 
Under Assumptions \ref{hyp2} and \ref{hyp3}, {\bf if} the following conditions hold:
\begin{enumerate}
\item $q^*(x_k,z_k)=1$  
\item $z_{k+1}<W(x_{k+1})$
\item $\alpha\ge 2N\bar L/(1-\gamma)$
\end{enumerate} 
{\bf then} the following inequality holds:
\begin{equation}
J^*(x_{k+1},z_{k+1})\le J^*(x_k,z_k)-z_kQ(x_{k+1})\label{fgdrzwz} 
\end{equation}   
\end{lemma}
{\sc Proof}. Since $q^*(x_k,z_k)=1$ and by virtue of the receding-horizon implementation, one has:
\begin{equation}
J^*(x_k,z_k)=z_kL(x_{k+1},\bm u^*_1(x_k,u_k))+\alpha W(x_{k+1}) 
\end{equation} 
therefore, since $L(x,u)\ge Q(x)$, the last inequality implies that:
\begin{equation}
\alpha W(x_{k+1})\le J^*(x_k,z_k)-z_kQ(x_{k+1})\label{GFTTT} 
\end{equation} 
On the other hand, since the conditions of Corollary \ref{cor1} are satisfied at instant $k+1$, inequality (\ref{nbnbnhjgkh}) holds for $x_{k+1}$ and $z_{k+1}$, namely:
\begin{equation*}
J^*(x_{k+1},z_{k+1})\le \left[\dfrac{1+\gamma}{2}\right]\alpha W(x_{k+1}) \le \alpha W(x_{k+1})
\end{equation*}   
Combining this last inequality with (\ref{GFTTT}) gives (\ref{fgdrzwz}).  $\hfill \Box$ \\ \ \\   
Lemma \ref{lem2} and \ref{lem3} enables to establish the following corollary:
\begin{corollary} \label{cor2} 
Under Assumptions \ref{hyp2} and \ref{hyp3}, {\bf If} the a penalty $\alpha\ge 2N\bar L/(1-\gamma)$ is used, then for all initial $z_0>0$, the set $\{x=0\}$  is an accumulation set for the closed-loop dynamic system. Namely, there is a subsequence of $\{x_k\}_{k\ge 0}$ that converges to $0$.
\end{corollary}
{\sc Proof}. Let us adopt the following notation:
\begin{equation}
e_k=W(x_k)-z_k \label{defdeek} 
\end{equation} 
then the updating rule (\ref{defdezplus}) can be rewritten for clarity using $e_k$ as follows:
\begin{align}
z_{k+1}=\left\{ 
\begin{array}{ll}
 z_k& \mbox{\rm if $e_k>0$}  \\
\beta z_k& \mbox{if $e_k\le 0$}
\end{array}
\right. \label{defdezpluse} 
\end{align}  
Combining Lemmas \ref{lem2} and \ref{lem3} (the conditions of which are satisfied), one can write that:
\begin{align}
&\Bigl\{e_k>0\ \mbox{\rm and}\  e_{k+1}>0\Bigr\} \Rightarrow \label{implication} \\ &J^*(x_{k+1},z_{k+1})\le J^*(x_k,z_k)-z_{k}Q(x_{k+1})\nonumber 
\end{align} 
Let us denote by $\mathbb K_\le$ the set of instants such that $e_k\le 0$, more precisely:
\begin{equation}
\mathbb K_\le:=\Bigl\{\kappa_1,\kappa_2,\dots\Bigr\}\quad \mbox{\rm where}\quad  e_{\kappa_j}\le 0
\end{equation} 
Two situations have to be distinguished: \ \\ \ \\  In the first, the set $\mathbb K_\le$ is finite with cardinality $\sigma=card(\mathbb K_\le)$ while in the second, the set $\mathbb K_\le$ is infinite. \ \\ \ \\ 
\underline{\sl Case where $\mathbb K_\le$ is finite with $\sigma=card(\mathbb K_\le)$}
In this case, one has:
\begin{equation}
(\forall k>\kappa_\sigma)\quad e_k>0
\end{equation} 
and therefore, by virtue of (\ref{defdezpluse}), it is possible to write:
\begin{equation}
(\forall k>\kappa_\sigma)\quad z_k=:z_\infty:=\beta^\sigma z_0>0 \label{hgYTFrz} 
\end{equation} 
and injecting this in (\ref{implication}) enables to write:
\begin{align}
&(\forall k>\kappa_\sigma),\\
&J^*(x_{k+1},z_{k+1})\le J^*(x_k,z_k)-z_\infty Q(x_{k+1}) \label{ENFGRFR} 
\end{align}  
which obviously proves that the sequence $\{Q(x_k)\}_{k> \kappa_\sigma}$ converges to $0$ and so does the sequence $\{x_k\}_{k\ge 0}$ \ \\ \ \\ 
\underline{\sl Case where $\mathbb K_\le$ is infinite}\ \\ \ \\ 
In this case, by definition of the updating rule where the second branch is visited an infinite number of times, one obviously has:
\begin{equation}
\lim_{k\rightarrow\infty} z_k=0
\end{equation} 
and by definition of the instant $e_{\kappa_j}$, it comes that:
\begin{equation}
\lim_{j\rightarrow \infty} W(x_{\kappa_j})\le \lim_{j\rightarrow \infty}z_{\kappa_j}=0 \label{HHH7} 
\end{equation} 
which shows that there is a partial sequence of $\{x_k\}_{k\ge 0}$ that converges to $0$. $\hfill \Box$ \ \\ \ \\ 
Corollary \ref{cor2} proves that the trajectory of the state visits regularly an always smaller neighborhood of the targeted state $x=0$ at an increasing infinite instants of time. It remains to analyze the asymptotic behavior of excursion of the state trajectory between these instants. In order to do this, a local assumption is needed regarding the property of the system in an arbitrary small neighborhood of the origin:
\begin{hyp}\label{hyp4} 
The origin is locally $N$-step stabilizable. More precisely, there is a local neighborhood $\mathcal V$ of the origin such that for all $x_0\in \mathcal V$, there is an admissible control sequence $\bm u$ such that $\bm \Pi_c\left[x^{\bm u}_N(x_0)\right]=0$ [where $\Pi_c(x)$ is the controllable substate corresponding to $x$ of the linearized system around the origin]. Moreover, the corresponding state trajectory is entirely contained in $\mathbb G$.
\end{hyp}
\begin{rem}
It is worth underlying that this assumption is much less stringent than the $N$-reachability assumption (in the large) as used in the early formulations of stable NMPC \cite{Keerthi1988,Mayne1990}. Indeed, the assumption used here is only local and imposes the reachability-in-$N$-step assumption only on a small neighborhood of the targeted state that can be as small as necessary.  
\end{rem}
We now have all we need to state the main result of this contribution:
\begin{proposition} \label{prop1} {\bf Assume that}:
\begin{enumerate}
\item Assumptions \ref{hyp2}-\ref{hyp4} are satisfied. 
\item The penalty $\alpha$ involved in the cost function (\ref{costdef1}) is such that
\begin{equation}
\alpha \ge \dfrac{2N\bar L}{1-\gamma} \label{boundinfonalpha} 
\end{equation}  
\end{enumerate} 
{\bf Then} $x=0$ is asymptotically stable for the closed-loop associated to the MPC law defined by (\ref{costdef1})-(\ref{costdef3}) for all initial states $(x,z)$ such that $x\in \mathbb G$ and $z>0$.
\end{proposition}
{\sc Proof}. We shall characterize the behavior of the state trajectory over time. To do this, let us divide the time instants into two sets, namely:
\begin{align}
\mathbb K_\le&:= \Bigl\{k\in \mathbb N\ \vert\ e_k:=W(x_k)-z_k\le 0\Bigr\} \label{defdeKle}\\ 
\mathbb K_>&:= \Bigl\{k\in \mathbb N\ \vert\ e_k:=W(x_k)-z_k> 0\Bigr\} \label{defdeKgt}
\end{align} 
Note that we already encountered $\mathbb K_\le$ in the proof of Corollary \ref{cor2}. The behavior of the state trajectory over the set $\mathbb K_\le$ is easy to characterize since by virtue of (\ref{hgYTFrz}), one has:
\begin{equation}
(\forall k\in \mathbb K_\le)\qquad W(x_k)\le \left[\beta^{m_k-1}\right]z_0
\end{equation}  
where for all $k\in \mathbb K_\le$, the integer $m_k$ denotes the order of $k$ in $\mathbb K_\le$.\ \\ \ \\ 
As for $\mathbb K_>$, the two cases regarding whether $\mathbb K_\le$ is finite or not have to be distinguished. \ \\ \ \\ 
\underline{\sl Case where $\mathbb K_\le$ is finite with cardinality $\sigma$}.\ \\ 
In this case, the proof of Corollary \ref{cor2} already showed that the inequality (\ref{ENFGRFR}) becomes satisfied after a finite number of steps $k=\kappa_\sigma+1$ and therefore, the behavior of the closed-loop trajectory is such that:
\begin{equation}
\sum_{k=\kappa_\sigma+2}^\infty Q(x_k)\le \dfrac{1}{z_\infty}\left[J^*(x_{\kappa_\sigma+1},z_\infty)\right]<\infty
\end{equation} 
where $z_\infty:=\beta^\sigma z_0>0$. 
  \ \\ \ \\ 
\underline{\sl Case where $\mathbb K_\le$ is infinite}.\ \\ \ \\ 
Note that in this case, thanks to (\ref{HHH7}) we have the characterization of the behavior over $\mathbb K_\le$. As for the behavior over instants in $\mathbb K_>$, insight can be obtained by observing that between any two successive instants $\kappa_j, \kappa_{j+1}\in \mathbb K_\le$, a constant and non vanishing $z_{\kappa_j+1}=C_j\neq 0$ applies and therefore, one obtains the same optimal solution (and hence the same state trajectory) if the cost function is divided by $C_j$ to get the modified cost
\begin{equation}
\Phi(x,\bm u,q)+\left[\dfrac{\alpha}{C_j}\right]\underline W(x,\bm u,q)
\end{equation} 
Moreover, since we know that $\ell^*=q^*$, the second term can simply be replaced by a penalty on the final value to get the following modified cost function:
\begin{equation}
\Phi(x,\bm u,q)+\left[\dfrac{\alpha}{C_j}\right]W(x,\bm u,q)
\end{equation} 
Now since $\lim_{j\rightarrow\infty} C_j=0$, the corresponding MPC formulation behaves asymptotically as a MPC formulation with final equality constraints on the controllable sub-state. We know that such formulation under Assumption \ref{hyp4} (that becomes true for sufficiently high $j$ for which $x_{\kappa_j}\in \mathcal V$) and the positive definiteness of $L$ used in $\Phi$ leads to well qualified and stable behavior over the interval $[\kappa_{j}+1,\kappa_{j+1}]$ \cite{Mayne2000}. This clearly ends the proof. $\hfill \Box$ \ \\ \ \\ 
Note that the condition (\ref{boundinfonalpha}) of Proposition \ref{prop1} is a quantified realization of the stabilizing role of terminal penalty in the context of absence of terminal constraint as suggested by \cite{Grune2010} (See the discussion of Section 8.2 regarding this issue). 
\section{Implementation Issues} \label{secimplementation} 
\noindent In this section, it is shown that the formulation (\ref{costdef1})-(\ref{costdef3}) which involves the integer variable $q$ (representing the free prediction horizon) can be replaced by a new formulation in which a predicted sequence $\bm u$ and a prediction horizon $q$ can be obtained by a two-stage algorithm in which, each step involves a given  prediction horizon ($\le N$) which is not a decision variable. Moreover, the so computed sequence, when implemented in a receding-horizon way, induces the convergence property established so far. \ \\ \ \\ 
More precisely, the two steps are defined as follows:
\begin{enumerate}
\item First, problem (\ref{costdef1})-(\ref{costdef3}) is solved for $z=0$ and with the additional constraints $q=N$. More precisely, the following {\bf fixed-horizon} optimization problem is solved:
\begin{align}
\min_{\bm u\in \mathbb U^N} \Bigl[\underline W(x,\bm u,N)\Bigr] \quad \vert \quad   \bm x^{\bm u}_\ell(x)\in \mathbb G\quad \forall\ell\le N\label{costdef12}
\end{align} 
to get the index $\ell^{\ddag}_N\in \{1,\dots,N\}$ of the instant where the maximum contraction occurs. Note that this is a standard optimization problem in the continuous variable $\bm u$.
\item  Using the resulting index $\ell^{\ddag}_N$, the following fixed-horizon optimization problem is solved in which, the original stage cost is re-introduced with the same penalty $z$:
\begin{align}
&\bm u^\ddag(x,z)\leftarrow \nonumber \\
&\min_{\bm u\in \mathbb U^{\ell^{\ddag}_N}} \left[J^{(x,z)}(\bm u,\ell^{\ddag}_N)\right]\quad \vert \quad   \bm x^{\bm u}_\ell(x)\in \mathbb G\quad \forall\ell\le \ell^{\ddag}_N \label{costdef23}
\end{align} 
which is nothing but (\ref{costdef1})-(\ref{costdef3}) in which $q=\ell^{\ddag}_N$ is used to get rid of the integer decision variable $q$. Again, this yields a standard optimization problem in the continuous variable $\bm u$.
\end{enumerate} 
Note that both problems inherit the advantages of the contraction-based formulation of the preceding section, namely the short prediction horizon and the absence of stability-related terminal constraints. Moreover, one has the following convergence result:
\begin{proposition}\label{proptwostage} 
{\bf If the following conditions hold}:
\begin{enumerate}
\item Assumptions \ref{hyp2}-\ref{hyp4} are satisfied. 
\item The penalty $\alpha$ involved in the definition of the cost function $J^{(x,z)}$ used in (\ref{costdef23}) satisfies  (\ref{boundinfonalpha}), 
\end{enumerate} 
{\bf Then} $x=0$ is asymptotically stable for the closed-loop associated to the MPC law defined by $\bm u^\ddag_1(\cdot,\cdot)$ for all initial state $(x,z)$ such that $x\in \mathbb G$ and $z>0$.
\end{proposition}
{\sc Proof}. This comes from the fact that at each decision instant, the first stage of the procedure computes the prediction horizon $\ell^*_N$ for which the maximum contraction sequence remains available for the second stage. This means that each time a candidate sequence that realizes the maximum contraction scenario is invoked as a candidate sequence in the proof of convergence for the original formulation, the argument is still valid for the modified formulation and the sequence of arguments can be repeated to derive the convergence result in the modified case. $\hfill \Box$
\section{Illustrative Example} \label{secexample}
\noindent Let us consider the discrete-time version of the nonholonomic system given by:
\begin{align}
x_1^+&=x_1+u_1 \label{syst1}\\
x_2^+&=x_2+u_2 \label{syst2}\\
x_3^+&=x_3+x_1u_2 \label{syst3}
\end{align} 
where the control vector $u$ is saturated according to:
\begin{equation}
\vert u_i\vert \le \bar u_i\qquad i\in \{1,2\}
\end{equation} 
while the state $x$ is constrained by:
\begin{equation}
x_1\in [-\rho,+\rho]\qquad \vert x_i\vert \le b\quad i=2,3
\end{equation} 
Let us consider the constraint set given by the constraints map:
\begin{equation}
g(x):= \begin{pmatrix}
\vert x_1\vert - \rho\cr 
x_2^2+x_3^3-b^2
\end{pmatrix} 
\end{equation} 
then the following result holds regarding the satisfaction of Assumption \ref{hyp2}:
\begin{proposition} \label{propex} 
If $\bar u_1\ge 2\rho$ and $\bar u_2=\mu b$ then Assumption \ref{hyp2} holds for any $b$, any $N\ge 3$ and any $\gamma\in (0,1)$ satisfying:
\begin{equation}
\gamma\ge 1-\mu \label{condgammaexample} 
\end{equation} 
Moreover the contraction map is given by $W(x)=\|x\|^2_2$. 
\end{proposition}
{\sc Proof}. See Appendix \ref{appendixproof1}. \ \\ \ \\  
In what follows $\bar u_1=2\rho$ and $\bar u_2=\mu b$ are used. Note that $\mu$ is the ratio between the bound $\bar u_2$ on $u_2$ and the radius of the admissible region in $(x_2,x_3)$. The result of Proposition \ref{propex} states that this ratio can be as small as desired, the contraction property still hold for $N=3$ and $\gamma=1-\mu$. This obviously shows that while the prediction horizon needed by standard terminal region MPC formulations would increase indefinitely, the contractive-based formulation would still need only $N=3$ to have its requirement satisfied. \ \\ \ \\ 
In the simulation, the following two different stage costs are considered:
\begin{align}
L_1(x,u)&=\|x\|^2+0.1\|u\|^2 \label{defdeL1}\\
L_2(x,u)&=0.01 x_1^2+x_2^2+100(x_2-x_3)^2+0.1\|u\|^2 \label{defdeL2} 
\end{align} 
which corresponds to the following upper bounds to be used in the computation of the convenient terminal penalty $\alpha$ involved in Propositions \ref{prop1} and \ref{proptwostage}:
\begin{align}
\bar L_1&:=\rho^2+2b^2+0.1\left[4\rho^2+(\mu b)^2\right] \label{defdeeLbar1}\\
\bar L_2&:=0.01\rho^2+401b^2+0.1\left[4\rho^2+(\mu b)^2\right] \label{defdeeLbar2} 
\end{align}  
Depending on the stage cost being used, $\alpha$ is taken equal to the minimal value required by the inequality (\ref{boundinfonalpha}) of Proposition \ref{prop1}.
\begin{figure}
\begin{center}
\includegraphics[width=0.48\textwidth]{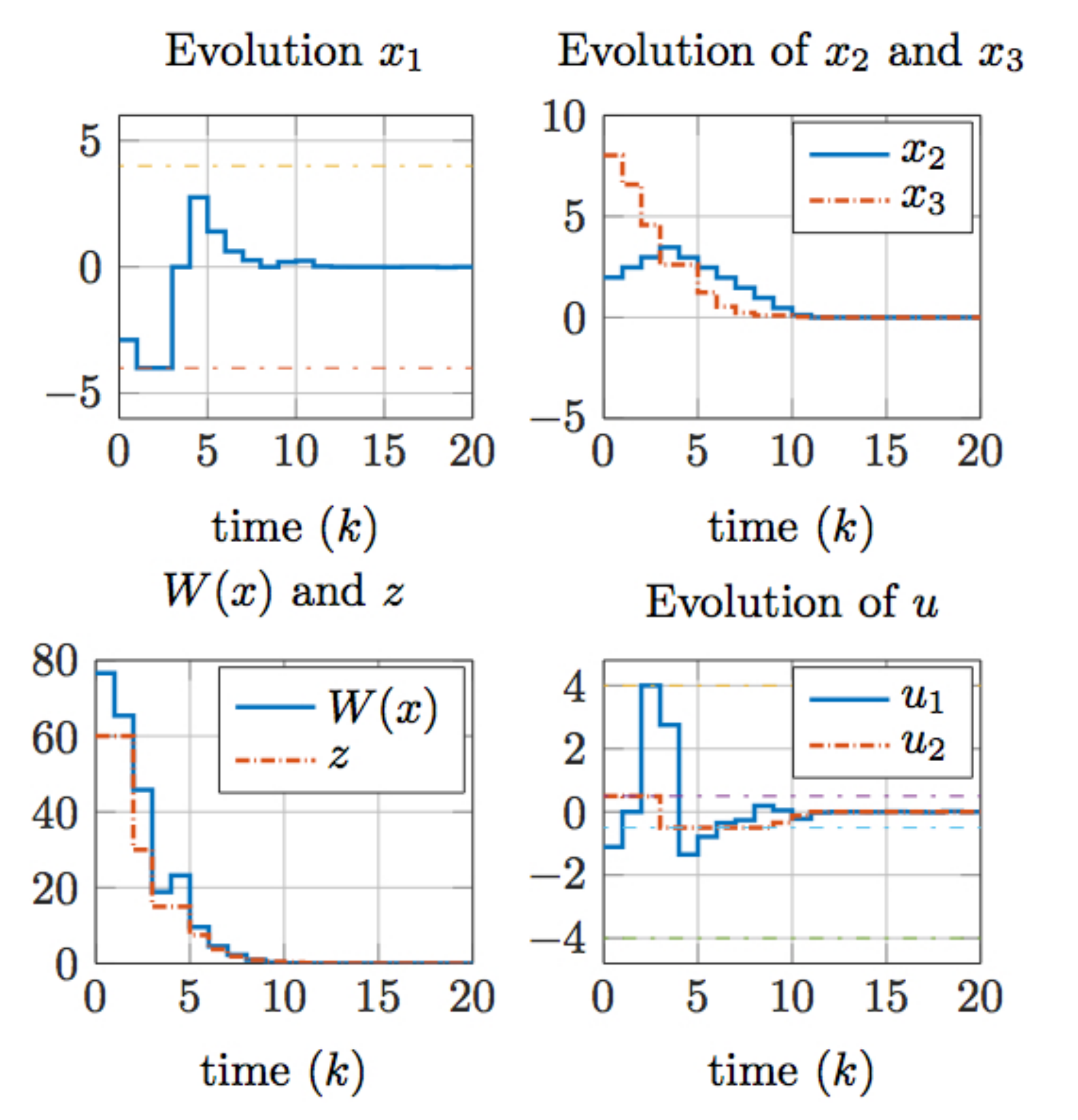} 
\end{center}
\caption{Evolution of the closed-loop system for $N=3$ and the stage cost $L_1(x,u)$ given by (\ref{defdeL1}).} \label{fig1}  
\end{figure}
\begin{figure}
\begin{center}
\includegraphics[width=0.48\textwidth]{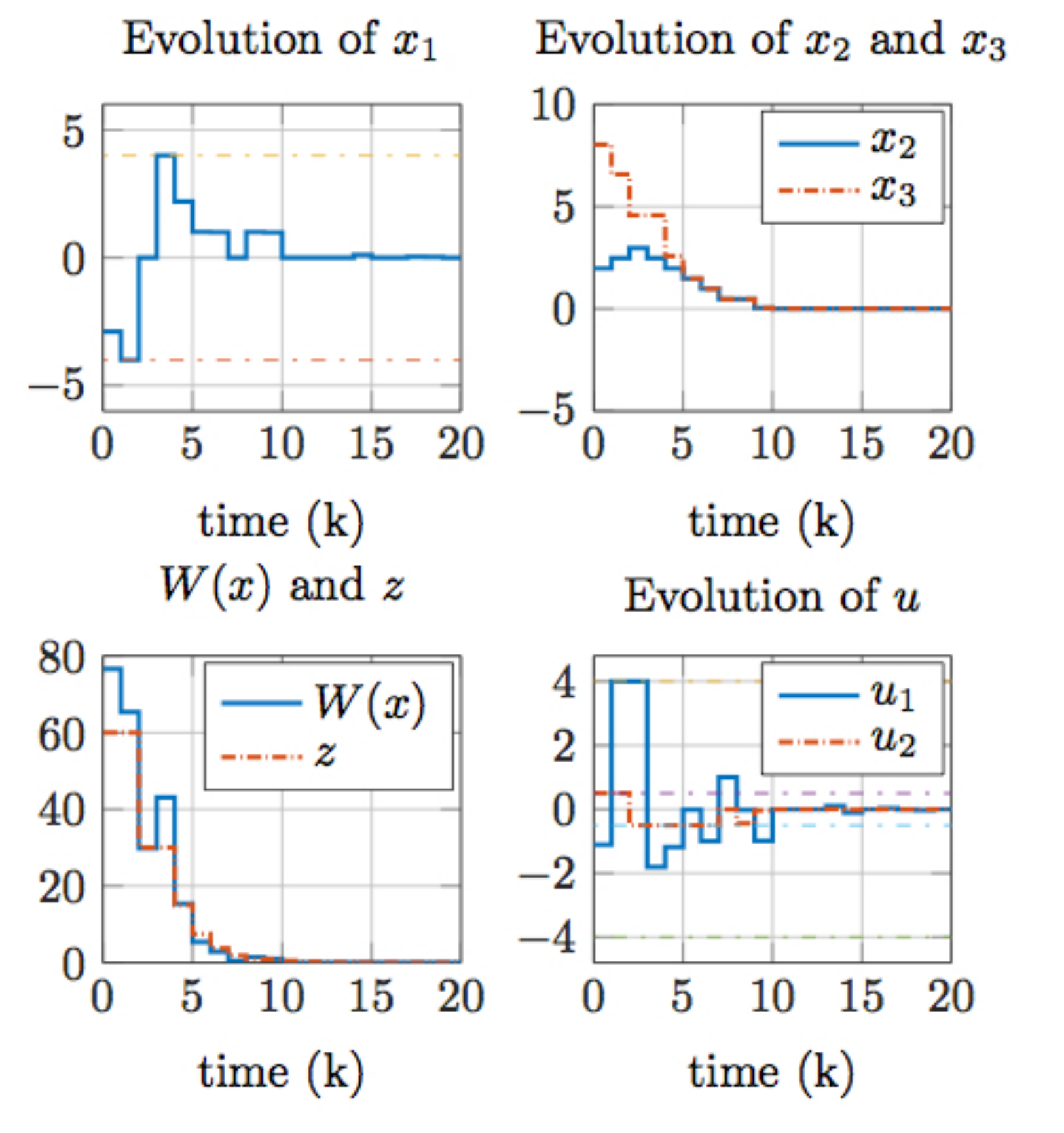} 
\end{center}
\caption{Evolution of the closed-loop system for $N=5$ and the stage cost $L_2(x,u)$ given by (\ref{defdeL1}).} \label{fig2}  
\end{figure}
\begin{figure}
\begin{center}
\includegraphics[width=0.48\textwidth]{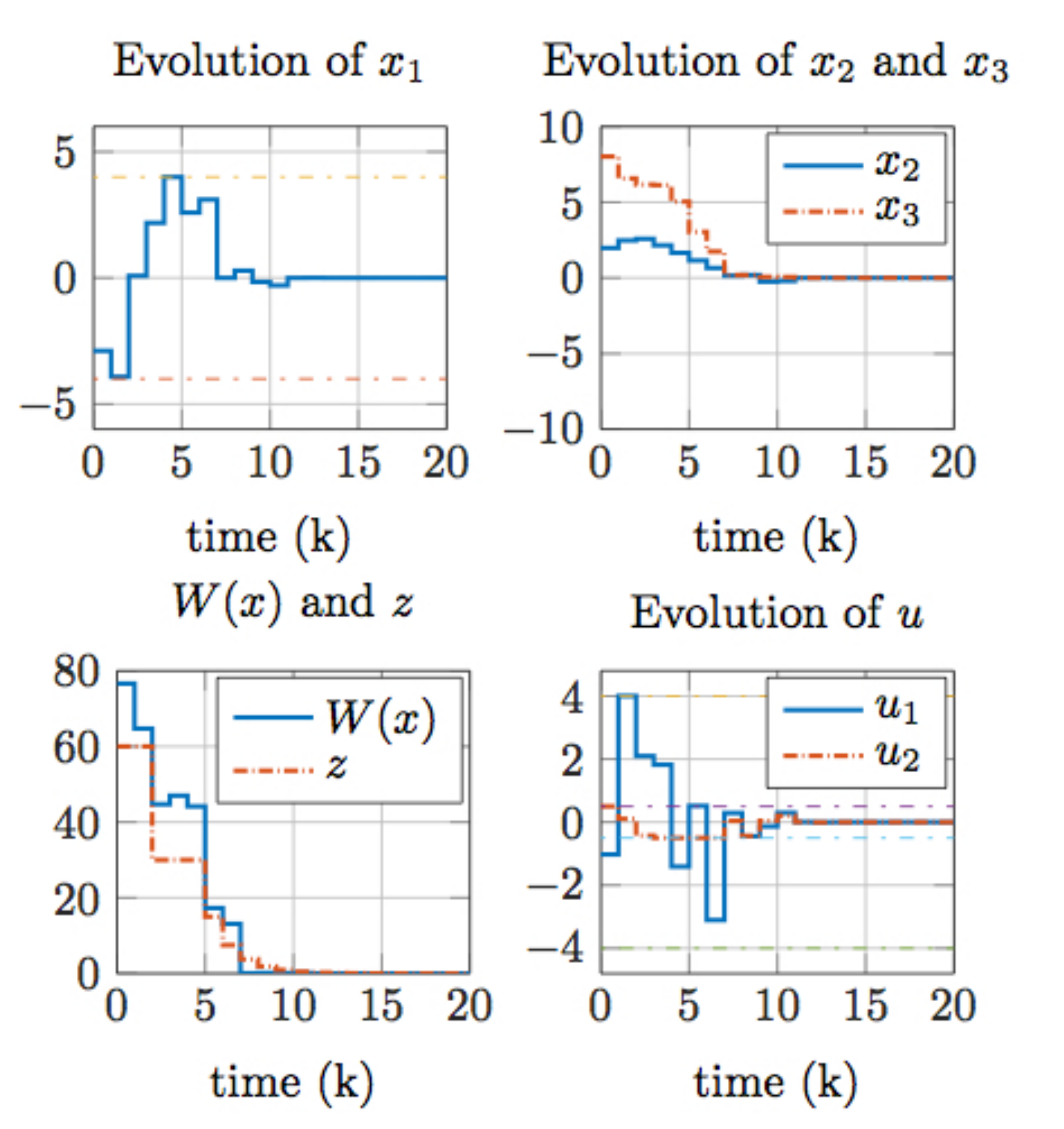} 
\end{center}
\caption{Evolution of the closed-loop system for $N=5$ and the stage cost $L_1(x,u)$ given by (\ref{defdeL1}).} \label{fig3}  
\end{figure}
In the following simulations, the following values are used:
\begin{equation}
\mu=0.05\quad,\quad b=10\quad,\quad \beta=0.5\quad,\quad \rho=4
\end{equation} 
Figure \ref{fig1} shows the closed-loop behavior when $N=3$ is used where the non monotonic decrease of the penalty function $W$ can be clearly observed. The fact that the stage cost does influence the behavior of the closed-loop system can be observed by comparing figures \ref{fig2} and \ref{fig3} where the stage costs $L_2$ and $L_1$ are successively used. Indeed, when $L_2$ is used, the difference $\vert x_2-x_3\vert$ is reduced when compared to the case where $L_1$ is used. This underline the advantage of the proposed formulation when compared to the previous contractive formulation \cite{alamir2007low} which needs the same function to be used in the penalty and the stage cost.
\section{Conclusion and future work} \label{secconc} 
\noindent In this paper, a new contraction-based NMPC formulation is proposed. The formulation uses the contraction property in the proof of the convergence of the closed-loop system but does not add any stability-related terminal constraint that involves the contraction property. The assumptions needed for the success of the formulation are rather standard and should be satisfied for short prediction horizons making the formulation adapted to situations where fast computation are necessary. \ \\ \ \\ 
Regarding future work, it can be conjectured that to any MPC provably stabilizable nonlinear system, one can associate a final constraint-free contractive formulation of the type proposed in the present contribution. Moreover, the associated penalty cost can be taken to be any positive function of the state although appropriate choice of this function can drastically reduce the prediction horizon. Proving this conjecture is an undergoing work. 

\bibliography{biblio_contractive}     
\bibliographystyle{plain}

\appendix 
\section{Proof of Proposition \ref{propex}} \label{appendixproof1} 
Assume that $x\in \mathbb G$. The proof would be obtained if one can find a sequence of length $N=3$ such that contraction occurs with $\gamma$ satisfying (\ref{condgammaexample}). We shall exhibit a sequence of the following form
\begin{equation}
\bm u= \Bigl\{ \begin{pmatrix}
x_1^*-x_1\cr 0
\end{pmatrix}, \begin{pmatrix}
0\cr u_2^*
\end{pmatrix}, \begin{pmatrix}
-x_1^*\cr 0
\end{pmatrix} \Bigr\} \label{defdelaseq} 
\end{equation} 
where $x_1^*$ and $u_2^*$ are to be found such that the intermediate visited states $\bm x^{\bm u}_\ell(x)$ for $\ell=1,2,3$ lie inside the admissible set $\mathbb G$ while the following contraction inequality holds:
\begin{equation}
W(\bm x^{\bm u}_N(x)):=\|\bm x^{\bm u}_3(x)\|_2^2\le \gamma \|x\|_2^2=:\gamma W(x)
\end{equation} 
which obviously implies (\ref{defdecontraction}) of Assumption \ref{hyp2}.   Let us introduce $z=(x_2,x_3)^T$ so that $x=(x_1,z^T)^T$. Note that by definition of the system dynamics and the definition of the control sequence (\ref{defdelaseq}), one has:
\begin{equation}
\bm x^{\bm u}_1(x)= \begin{pmatrix}
x_1^*\cr z
\end{pmatrix}\ ,\ \bm x^{\bm u}_2(x)=\begin{pmatrix}
x_1^*\cr z^*
\end{pmatrix}\ ,\ 
\bm x^{\bm u}_3(x)= \begin{pmatrix}
0\cr z^*
\end{pmatrix} 
\end{equation}  
where 
\begin{equation}
z^*:=z+ \begin{pmatrix}
1\cr x_1^*
\end{pmatrix}u_2^* \label{defdezstarttrtr} 
\end{equation} 
From these expressions, the following facts can be stated:
\begin{enumerate}
\item The trajectory lies entirely in $\mathbb G$ if $\vert x_1^*\vert\le 2$ and $\|z^*\|\le b$. 
\item $\gamma$-Contraction occurs if $\|z^*\|^2\le \gamma \|z\|^2$.
\item if both $x_1$ and $x_1^*$ are admissible then the control sequence (\ref{defdelaseq}) is admissible if $u_2^*$ is admissible (since $\bar u_1\ge 2\rho$ while $\vert x_1\vert$ and $\vert x_1^*\vert$ are both lower than $\rho$).
\end{enumerate} 
Note that the $\gamma$-contraction condition would imply that $\|z^*\|\le \gamma b$. Therefore, we have only to prove that there exists $x_1^*\in [-\rho,+\rho]$  and some $u_2^*\in [-\bar u_2,+\bar u_2]$ such that $\|z^*\|^2\le \gamma \|z\|^2$ with $\gamma$ satisfying (\ref{condgammaexample}). There are two cases to be distinguished: In the first case, $\bar u_2\ge b$ (that is $\mu\ge 1)$ in which case, for all $z$ such that $\|z\|\le b$, one can obviously find $x_1^*$ and $u_2^*$ such that $z^*=0$ making $\gamma=0$ a possible choice. In the more interesting second case where $\bar u_2:= \mu b<b$ for some $\mu\in (0,1)$, one can use the following argumentation: First of all, by symmetry w.r.t the origin, it is possible to consider only the case where $z_1\ge 0$ and $z_2\ge 0$. Moreover, the worst case regarding the contraction factor obviously occurs when $z$ lies on the boundary of the admissible region, that is $z=(\cos\phi,\sin\phi)^Tb$. Moreover, as everything is defined as a fraction of $b$, one can take $b=1$. To summarize, the contraction factor $\gamma$ can be given by the solution of the following optimization problem:
\begin{equation*}
\gamma^2 := \max_{\phi\in [0,\pi/2]}R(\phi):=\left[\begin{array}{l} \min_{v\ge 0} \| \begin{pmatrix}
\cos\phi\cr \sin\phi
\end{pmatrix}- \begin{pmatrix}
v_1\cr v_2
\end{pmatrix}\|^2\\
\mbox{\rm under} \\
v_1\in [0,\mu]\\
 v_2\in [0,\rho v_1]
\end{array}\right]
\end{equation*} 
Note that the constraints $v_1\in [0,\mu]$ and $v_2\in [0,\rho v_1]$ simply implements the conditions on $x_1^*$ and $u_2^*$ involved in the expression (\ref{defdezstarttrtr}) of $z^*$. \ \\ \ \\  
Now a deeper analysis of this problem shows that the maximum value of $R(\phi)$ occurs when $\phi=0$. This is simply because the bound of $v_1$ is $\rho$-time lower than the bound on $v_2$. Moreover, in this case, the solution is obviously given by $v^*:=(\mu,0)$ leading to the contraction factor $\gamma$ given by:
\begin{equation}
\gamma=1-\mu
\end{equation} 
This obviously ends the proof. $\hfill \Box$

\end{document}